\documentclass[%
superscriptaddress,
preprint,
amsmath,amssymb,
aps,
]{revtex4-1}

\usepackage{color}
\usepackage{graphicx}
\usepackage{dcolumn}
\usepackage{bm}
\usepackage{textcomp}

\begin {document}

\title{Non-Markovian effects of conformational fluctuations on the global diffusivity in Langevin equation with fluctuating diffusivity}

\author{Mutsumi Kimura}
\affiliation{%
  Department of Physics, Tokyo University of Science, Noda, Chiba 278-8510, Japan
}%

\author{Takuma Akimoto}
\email{takuma@rs.tus.ac.jp}
\affiliation{%
  Department of Physics, Tokyo University of Science, Noda, Chiba 278-8510, Japan
}%



\date{\today}

\begin{abstract}
Local diffusivity of a protein depends crucially on the conformation, and the conformational fluctuations are often  non-Markovian. 
Here, we investigate the Langevin equation with non-Markovian fluctuating diffusivity, where the fluctuating diffusivity is modeled by a generalized Langevin equation under 
a double-well potential. We find that non-Markovian fluctuating diffusivity affect the global diffusivity, i.e., the 
diffusion coefficient obtained by the long-time trajectories when the memory kernel in the generalized Langevin equation
is a power-law form. On the other hand,  the diffusion coefficient does not change when the memory kernel is exponential.  
We show that these non-Markovian effects  are the consequences of an everlasting effect of the initial condition on the stationary distribution 
in the generalized Langevin equation under a double-well potential due to long-term memory.
\end{abstract}

\maketitle


\section{Introduction}

A particle immersed in an aqueous solution exhibits a random motion known as the Brownian motion. 
The mean square displacement (MSD) of the random motion increases linearly with time $t$: $\langle x(t)^2 \rangle 
= 2Dt$, where $x(t)$ is the position of the particle with $x(0)=0$. The proportional 
constant $D$ characterizes the diffusivity, i.e., the diffusion coefficient. Diffusion is a consequence of collisions between  the particle and solvent molecules, where
the diffusing particle is much larger than the solvent molecules.
Therefore, the theory of diffusion is connecting microscopic dynamics and macroscopic behaviors such as the diffusion coefficient \cite{einstein1956investigations}. 
The Stokes-Einstein relation states that the diffusion coefficient of a spherical particle is described by
\begin{equation}
D= \frac{k_{\rm B}T}{6 \pi \eta R},
\end{equation}
where $k_{\rm B}, T, \eta$ and $R$ are the Boltzmann constant, temperature, viscosity, and the radius of the spherical particle \cite{stokes1851effect,einstein1956investigations}. 
Furthermore, a Stoke-Einstein-like relation is also considered to hold in lateral diffusion of a protein on the membrane, where the radius $R$ is replaced by a lateral radius of a protein \cite{saffman1975brownian, vaz1984lateral, javanainen2017diffusion}. 

Temporal fluctuations of the diffusion coefficient are intrinsic in the diffusion of proteins \cite{Manzo2015, yamamoto2017dynamic, Yamamoto2021}, supercooled liquids \cite{Yamamoto-Onuki-1998, Yamamoto-Onuki-1998a, Richert-2002}, and entangled polymers \cite{Doi-Edwards-book, Uneyama2012, Uneyama2015}. The reptation model represents the diffusion 
of entangled polymers \cite{Doi-Edwards-book}. The dynamic equation of the center-of-mass can be described by the end-to-end vector, which depends on the conformation 
and changes with time. Therefore, the local diffusivity (instantaneous diffusion coefficient) in the reptation model exhibits temporal fluctuations.  
Proteins in aqueous solutions diffuse and exhibit conformational 
fluctuations. Because the gyration radius changes with time in the diffusion of a protein, the local diffusivity also fluctuates \cite{Yamamoto2021}.  

Fluctuating diffusivity provides rich physical behaviors such as Brownian yet non-Gaussian diffusion \cite{wang2009anomalous, wang2012brownian}, anomalous diffusion and 
ergodicity breaking \cite{Manzo2015}. Diffusing diffusivity model is a theoretical model of Brownian yet non-Gaussian diffusion, which  exhibits 
an exponential distribution in the propagator \cite{Chubynsky2014, Miyaguchi2016, Chechkin2017}. The annealed transit time model exhibits ergodicity breaking and 
anomalous diffusion \cite{Massignan2014, akimoto2016distributional}. Furthermore, in Langevin equation with fluctuating diffusivity, it has been shown that 
trajectory-to-trajectory fluctuations of the time-averaged MSD  are intrinsic 
when the measurement time is not greater than the largest relaxation time \cite{Uneyama2012, Uneyama2015, Akimoto2016PRE}. 

Temporal protein conformational fluctuations are often non-Markov \cite{yang2003protein, Min2005, Yamamoto2014b, ayaz2021non, Yamamoto2021}. In particular, temporal fluctuations of
 the gyration radius exhibit $1/f$ noise \cite{Yamamoto2021}. Because the local diffusivity is related to the gyration radius, it is expected that 
 fluctuating diffusivity is also described by a non-Markov process. Therefore, it is important to clarify the effect of non-Markovian fluctuations of instantaneous diffusion coefficient. 
 Several non-Markov diffusing diffusivity models were proposed to investigate the non-Gaussianity of the propagator and anomalous diffusion \cite{Wang_2020, Miyaguchi2022}. 
However, it is still an open problem how the non-Markovian fluctuating diffusivities affect the global diffusivity, i.e., the diffusion coefficient obtained by the long-time trajectories. 

Here, we investigate the Langevin equation with non-Markovian fluctuating diffusivity to unravel the effects of non-Markovian fluctuating diffusivity on the global diffusivity.
In particular, we model the dynamics of the gyration radius of a protein by the generalized Langevin equation under a double-well potential 
and assume the Stokes-Einstein-like relation. There are two minima in the double-well potential, which correspond to the folding and unfolding states of a protein. 
We consider power-law and exponential memory kernels to clarify the effects of the memory kernels on the diffusion coefficient. 

\section{Langevin equation with non-Markovian fluctuating diffusivity}

\begin{figure}
\includegraphics[width=.9\linewidth, angle=0]{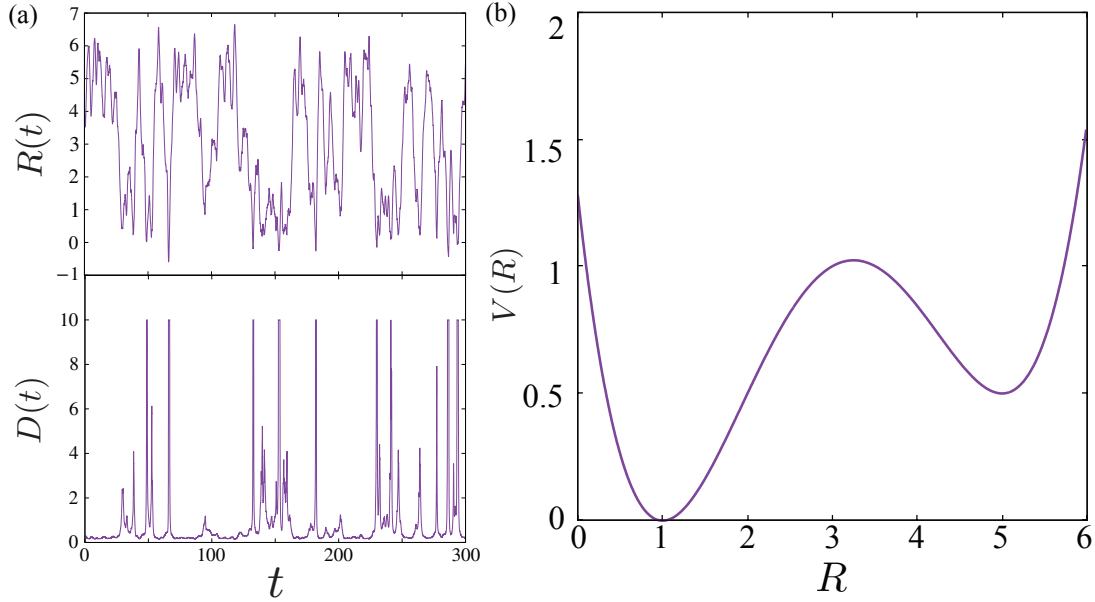}
\caption{(a) trajectory $R(t)$, and the corresponding fluctuating diffusivity $D(t)$. (b) Double well potential, where the parameters are  $a=0.046875, b=-0.578125, c=2.296875, d=-3.046875$, and $e=1.28125$. There are two minima  in the double-well potential, i.e., $R=1$ and 5.}
\label{trajectory}
\end{figure}

We consider a Langevin equation with non-Markovian fluctuating diffusivity as a model of diffusion of a protein. 
The dynamic equation of  the Langevin equation with fluctuating diffusivity (LEFD) is described as
\begin{equation}
\frac{dx}{dt} = \sqrt{2D(t)} \xi (t), 
\end{equation}
where $x(t)$ is a position (center-of-mass) of a molecule, $D(t)$ is the fluctuating diffusion constant, and $\xi(t)$ is a white Gaussian noise with 
$\langle \xi (t) \xi(t')\rangle = \delta (t-t')$. 
We assume the Stokes-Einstein-like relation for the fluctuating diffusivity $D(t)$, i.e., 
\begin{equation}
D(t) = \frac{1}{R(t)},
\end{equation}
where $R(t)$ represents the gyration radius of the molecule. We assume that the dynamic equation for $R(t)$ is described by the generalized Langevin equation (GLE):
\begin{equation}
m \frac{d^2 R}{dt^2} = - \int_0^t \eta (t-t') v_R(t') dt' - \frac{\partial V(R)}{\partial R} +   \xi_R (t), 
\label{GLE}
\end{equation}
where $\eta(t)$ is the memory kernel, $v_R(t)=dR(t)/dt$, $V(R)$ is a potential, and $\xi_R(t)$ is a noise which satisfies the fluctuation-dissipation relation:
\begin{equation}
\langle \xi_R(t) \xi_R(t') \rangle =  k_{\rm B} T \eta (t-t').
\label{fd relation}
\end{equation}
In what follows, we consider a double-well potential:
\begin{equation}
V(R) = aR^4 + bR^3 + cR^2 + d R +e.
\end{equation}
There are two minima in the double-well potential (see Fig.~\ref{trajectory}). The parameters in the GLE are set to be $m=1$ and $k_{\rm B} T=1$. 
To prevent a negative diffusivity and an extreme large diffusivity, we assume $D(t)=10$ for $R(t)<0.1$. 
Figure~\ref{trajectory} shows a trajectory of $R(t)$ under a double well potential and the corresponding fluctuating diffusivity $D(t)$.

We consider  exponential and power-law kernels in the GLE.  The exponential kernel is given by
\begin{equation}
\eta (t) = \eta_0 \lambda e^{-\lambda |t|},
\end{equation}
where $\eta_0$ is the intensity and $\lambda$ is the inverse relaxation time. Here, we derive Markovian Langevin equations from the GLE with an 
exponential kernel, which is used in numerical simulations. First, we consider an Ornstein-Uhlenbeck process $S(t)$ \cite{Uhlenbeck1930}, where the dynamics are 
described by the following equation:
\begin{equation}
\dot{S}(t) = -\lambda S(t) + \lambda \sqrt{2 k_B T \eta_0}\xi_0 (t),
\end{equation}
where $\xi_0 (t)$ is a white Gaussian noise, which satisfies $\langle \xi_0 (t)=0$ and $\langle \xi_0 (t)\xi_0 (t')\rangle =\delta(t-t')$. 
By a simple calculation, it is shown that the correlation function of $S(t)$ becomes an exponential decay \cite{kubo2012statistical}:
\begin{equation}
\langle S(t) S(t') \rangle =  k_B T \eta_0 \lambda e^{-\lambda |t-t'|}.
\end{equation}
This is the same as the fluctuation-dissipation relation of $\xi_R(t)$, i.e., Eq.~(\ref{fd relation}). Therefore, one can use $S(t)$ as noise $\xi_R(t)$. 
Furthermore, to eliminate the integral in Eq.~(\ref{GLE}),  we introduce an auxiliary variable $\zeta (t)$:
\begin{equation}
\sqrt{\lambda} \zeta (t) = - \int_0^t  \eta_0 \lambda e^{-\lambda |t-t'|} v_R(t')dt' +S(t).
\label{auxiliary var}
\end{equation}
Differentiating Eq.~(\ref{auxiliary var}), we have
\begin{equation}
\dot{\zeta} (t) = -   \eta_0 \sqrt{\lambda} v_R(t) -\lambda \zeta(t) + \sqrt{2\lambda  k_B T \eta_0}\xi_0 (t).
\label{auxiliary var2}
\end{equation}
As a result, we have the following Markovian Langevin equations:
\begin{align}
\dot{R}(t) &= v_R(t) ,\\
m\dot{v_R}(t) &= -\frac{\partial V(x)}{\partial x} + \sqrt{\lambda}\zeta (t) ,\\
\dot{\zeta} (t) &= - \eta_0 \sqrt{\lambda}v_R(t) -\lambda \zeta (t) + \sqrt{2\lambda   k_B T \eta_0}\xi (t).
\end{align}
We use these equations in numerical simulations. 

For a power-law memory kernel, we use  
\begin{equation}
\eta (t) = \frac{\eta_\alpha}{\Gamma (1-\alpha)} t^{-\alpha},
\end{equation}
where $\eta_\alpha$ is a constant, $\Gamma(\cdot)$ is the gamma function, and $\alpha$ is a power-law exponent. 
In numerical simulations of the GLE with a power-law memory kernel,  we use a Markov embedding 
method \cite{goychuk2009viscoelastic}, where we approximate the power-law memory kernel by a superposition of exponential 
functions:
\begin{equation}
\eta (t) = \sum_{i=0}^{N-1}\eta _i e^{-\nu _i t},
\end{equation}
where $\nu _i = \frac{\nu _0}{b^i}$ and $\eta _i = \frac{\eta _{\alpha}}{\Gamma (1-\alpha )}\frac{C_{\alpha}(b)\nu _0^{\alpha}}{b^{i\alpha}}$. 
Using auxiliary variables $u_i$, we have Markovian Langevin equations:  
\begin{align}
\dot{R}(t) &= v_R(t), \label{MGLE1}\\
m\dot{v}_R(t) &= -\frac{\partial V(x)}{\partial x} + \sum_{i=0}^{N-1} u_i(t),  \label{MGLE2} \\
\dot{u}_i (t) &= -\eta _iv_R(t) -\nu _{i}u_i(t) + \sqrt{2\nu _i \eta _i k_{\rm B} T}\xi _i(t),  \label{MGLE3}
\end{align}
where $\xi_i$ is a white Gaussian noise. In what follows, the parameters of the GLE with a power-law memory kernel are set to be 
$C_{\alpha}(b)=1, N=16, \nu_0=10^3, b=10, \eta_{\alpha}=1, k_{\rm B}=1, T=1$ and $m=1$.

\section{Non-Markovian effects of conformational fluctuation on the diffusivity}

The global diffusivity of the LEFD can be characterized by the asymptotic behavior of  the MSD. 
In many cases, the MSD of the LEFD exhibits normal diffusion \cite{Chubynsky2014, Miyaguchi2016, Chechkin2017, Uneyama2012, Uneyama2015, Akimoto2016PRE}: 
\begin{equation}
\langle x(t)^2 \rangle = 2 \langle D(t) \rangle t,
\end{equation}
where $\langle \cdot \rangle$ represents the ensemble average \cite{Uneyama2015}. If the stochastic process
$D(t)$ is stationary, the ensemble average $\langle D(t) \rangle$ does not depend on time $t$ and is given by 
the stationary distribution $\langle D \rangle_{\rm st}$. When the dynamics are Markovian, the stationary distribution is represented by the equilibrium distribution of $R$.
Therefore, the ensemble average is given by 
\begin{equation}
\langle D \rangle_{\rm st} = \int_{-\infty}^{\infty} D(R) P_{\rm eq}(R) dR,
\label{DC botzmann}
\end{equation}
where $P_{\rm eq}(R)$ is the equilibrium distribution of gyration radius $R$ and 
$D(R)$ is the instantaneous diffusion coefficient given that the gyration radius is $R$. 

When the dynamics of $R(t)$ are described by
 the Langevin equation, i.e., $\eta(t)=\delta(t)$, the equilibrium distribution is given by the Boltzmann distribution:
\begin{equation}
P_{\rm eq}(R) = N \exp \left( - \frac{V(R)}{k_{\rm B}T}\right),
\label{boltzmann dist}
\end{equation}
where $N$ is the normalization constant. Therefore, the equilibrium distribution of $R(t)$ determines the global diffusivity in the LEFD. 
However, when the dynamics of $R(t)$ are described by the GLE with a long-term memory, the equilibrium distribution may deviate from the 
Boltzmann distribution. Therefore, the global diffusivity may change due to non-Markovian fluctuating diffusivity even when fluctuating diffusivity 
is a stationary process.

\begin{figure}
\includegraphics[width=.9\linewidth, angle=0]{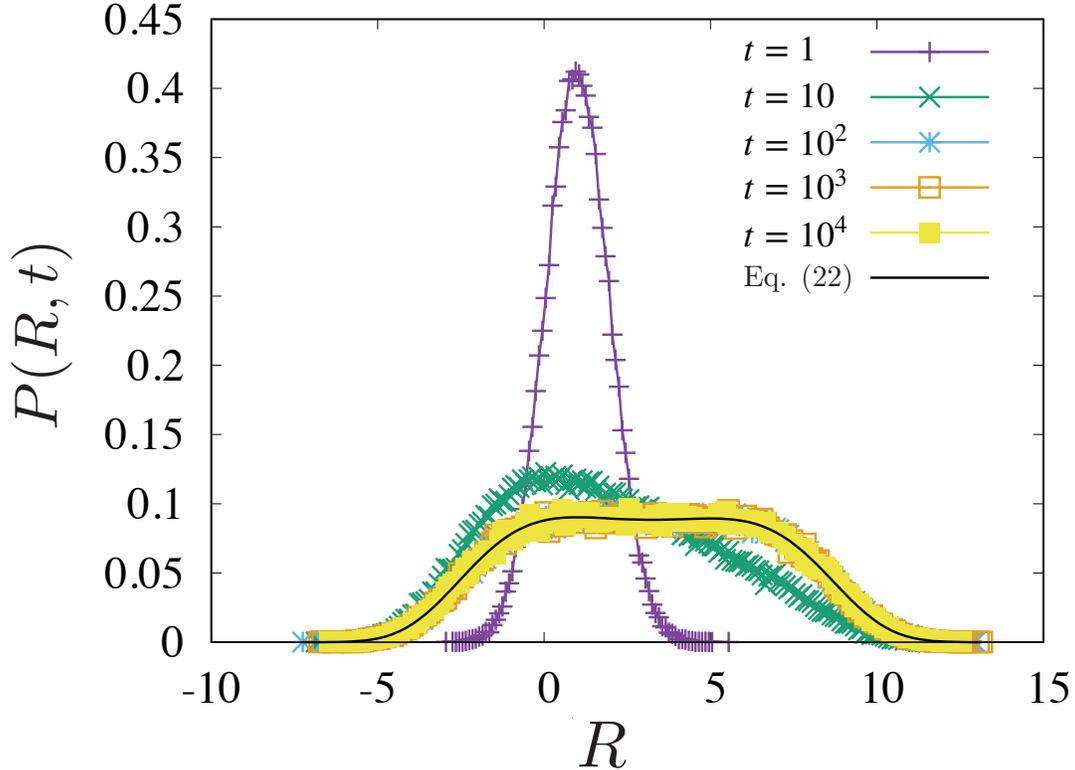}
\caption{Propagators of gyration radius $R(t)$ for different times $t$, where the dynamics of $R(t)$ are described by the GLE with an exponential memory kernel ($\lambda=1$), 
 the initial condition is $P(R,0) = \delta(R-1)$ and the parameters of the double-well potential are $a=0.000937,b=-0.011562,c=0.0045937,d=-0.060938$, and $0.025625$.
The solid line shows the equilibrium distribution, i.e., Eq.~(\ref{boltzmann dist}). }
\label{equilibrium}
\end{figure}


\subsection{exponential memory kernel}

When the memory kernel is exponential, the GLE can be described by the Markovian Langevin equations, i.e., Eqs.~(\ref{MGLE1}), (\ref{MGLE2}) and (\ref{MGLE3}). 
In other words, the equations are a three-dimensional Langevin equation.
Therefore, the Fokker-Planck equation can be described explicitly \cite{gardiner2009stochastic}. The stationary solution of the Fokker-Planck equation is obtained 
by the canonical distribution. Integrating the stationary distribution with respect to $v_R$ and $\zeta$, we have the stationary distribution for $R$.
In particular, the stationary distribution for $R$ is also given by the Boltzmann distribution, i.e., Eq.~(\ref{boltzmann dist}).
Figure~\ref{equilibrium} shows that the propagator $P(R,t)$ of $R(t)$ converges to the Boltzmann distribution in the long-time limit. 
Moreover, we confirm numerically that equilibrium distributions for different $\lambda$ do not depend on $\lambda$ and are described by the Boltzmann distribution. 
Therefore, the relaxation time $\lambda$ in the exponential memory kernel does not affect the equilibrium distribution. 

\begin{figure}
\includegraphics[width=.9\linewidth, angle=0]{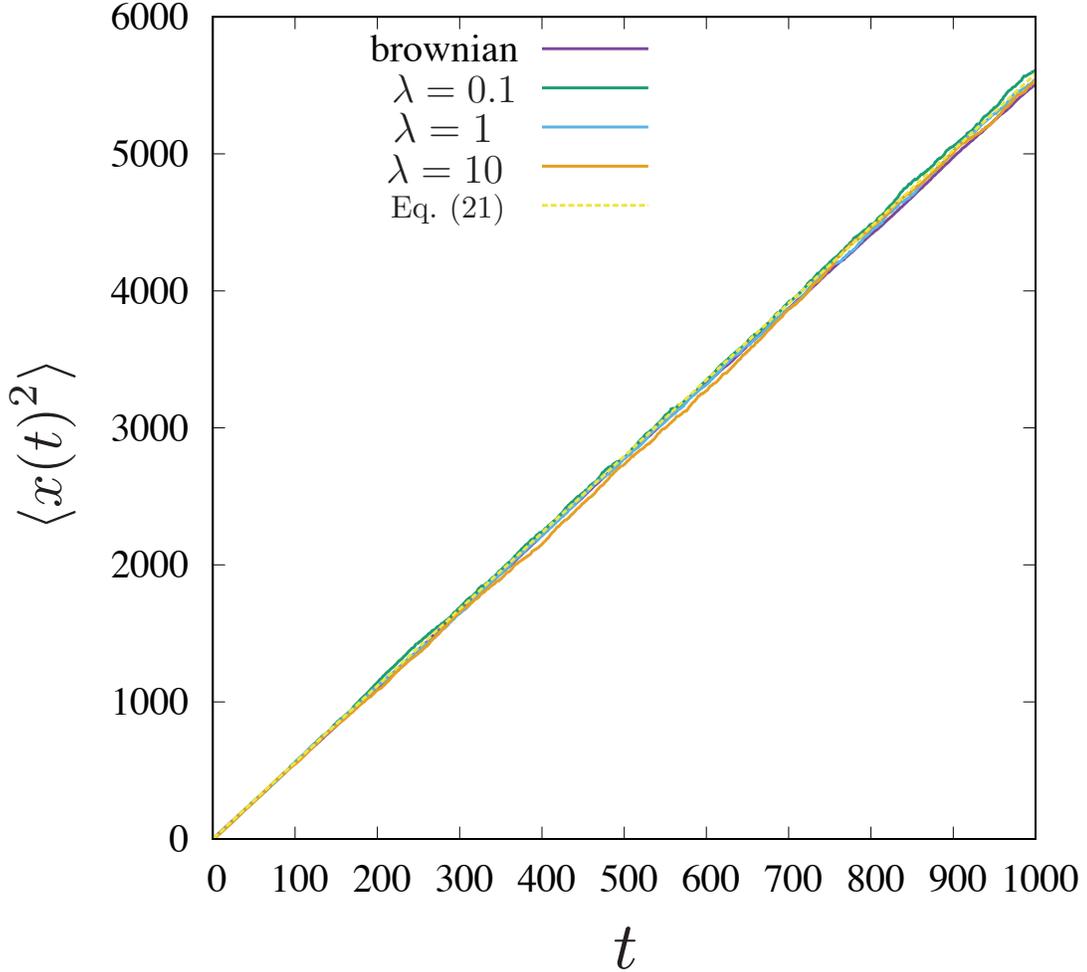}
\caption{Mean square displacement for the LEFD, where fluctuating diffusivity is obtained from the GLE with an exponential memory kernel 
($\lambda =$0.1, 1, and 10) and the parameters of the double-well potential are the same as those in Fig.~\ref{equilibrium}.
The dashed line represents Eq.~(\ref{DC botzmann}). The result for the case that $R(t)$ is the Brownian motion 
is also shown for reference.}
\label{msd exponential}
\end{figure}

The diffusion coefficient does not depend on $\lambda$ but is described by Eq.~(\ref{DC botzmann}) because the equilibrium distribution does not depend on $\lambda$.
Figure~\ref{msd exponential} shows the MSDs for different $\lambda$. All the MSDs exhibit normal diffusion, i.e., linear growth of $t$. 
The diffusion coefficient in the MSD is obtained by Eq.~(\ref{boltzmann dist}), i.e., the equilibrium distribution in the GLE of $R(t)$. 
As shown in Fig.~\ref{msd exponential}, the diffusion coefficients estimated by  Eq.~(\ref{boltzmann dist}) are in good agreement with numerical results. 
As a result, there is no non-Markovian effect on the global diffusivity when the memory kernel is exponential.

\subsection{power-law memory kernel}

\begin{figure}
\includegraphics[width=.9\linewidth, angle=0]{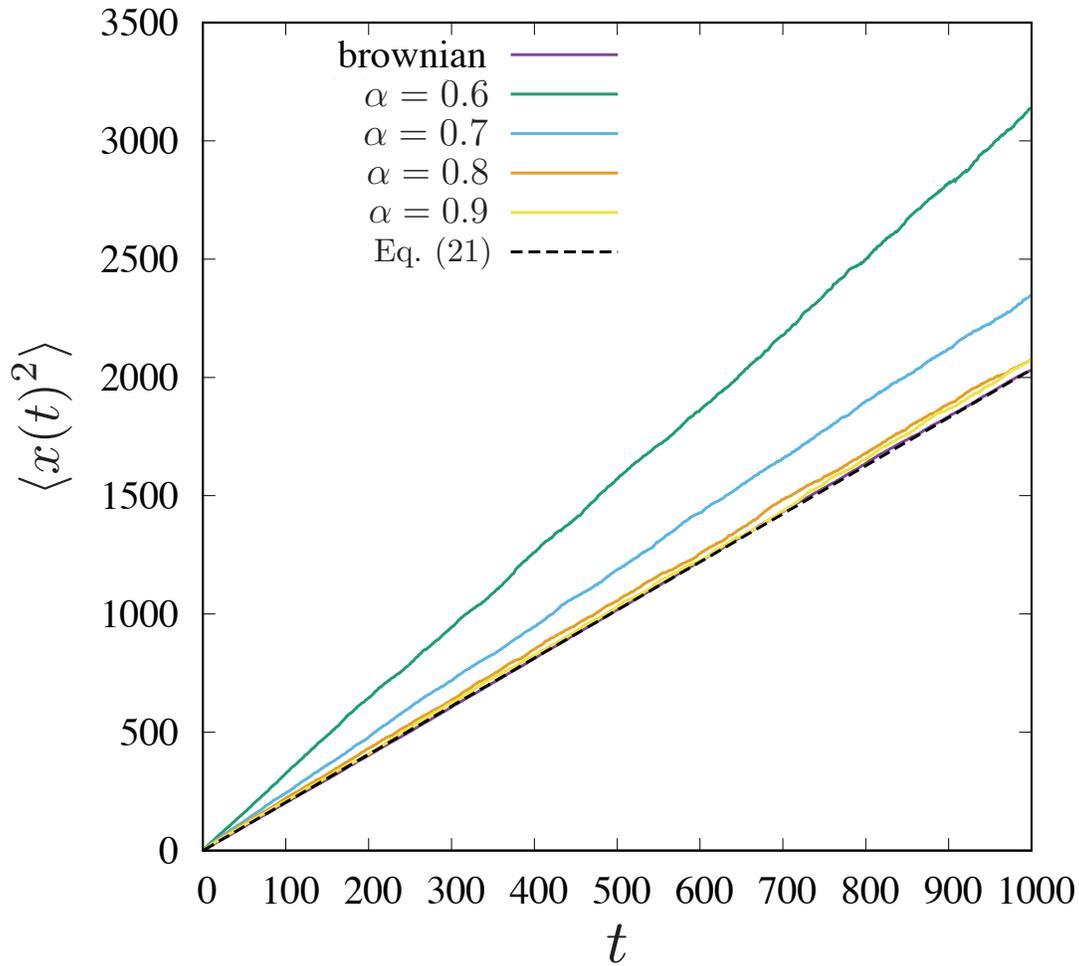}
\caption{Mean square displacement for the LEFD, where fluctuating diffusivity is obtained from the GLE with a power-law memory kernel 
($\alpha = $0.6, 0.7, 0.8, and 0.9) and the parameters of the double-well potential are the same as those in Fig.~\ref{trajectory}. The initial condition for $R(t)$ is located at 
the left minimum in the potential, i.e., $R(0)=1$. 
The dashed line represents Eq.~(\ref{DC botzmann}). The result for the case that $R(t)$ is the Brownian motion 
is also shown for reference.}
\label{msd power-law}
\end{figure}

Figure~\ref{msd power-law} shows the MSDs when the memory kernel is a power-law form. Even when the fluctuating diffusivity $D(t)$ has a long-term memory, 
all the MSDs exhibit normal diffusion.  This is because there exists a stationary distribution for $R(t)$, which means that $D(t)$ is also a stationary process. 
However,  the diffusion coefficients deviates from Eq.~(\ref{DC botzmann}) and depend on the power-law exponent $\alpha$. 
In particular, the diffusion coefficient is increased by the decrease of $\alpha$, which implies that the non-Markovian effect increases the diffusivity. 
Therefore, we find that there is a clear non-Markovian effect on the global diffusivity when the memory kernel is a power-law form.

\begin{figure}
\includegraphics[width=.9\linewidth, angle=0]{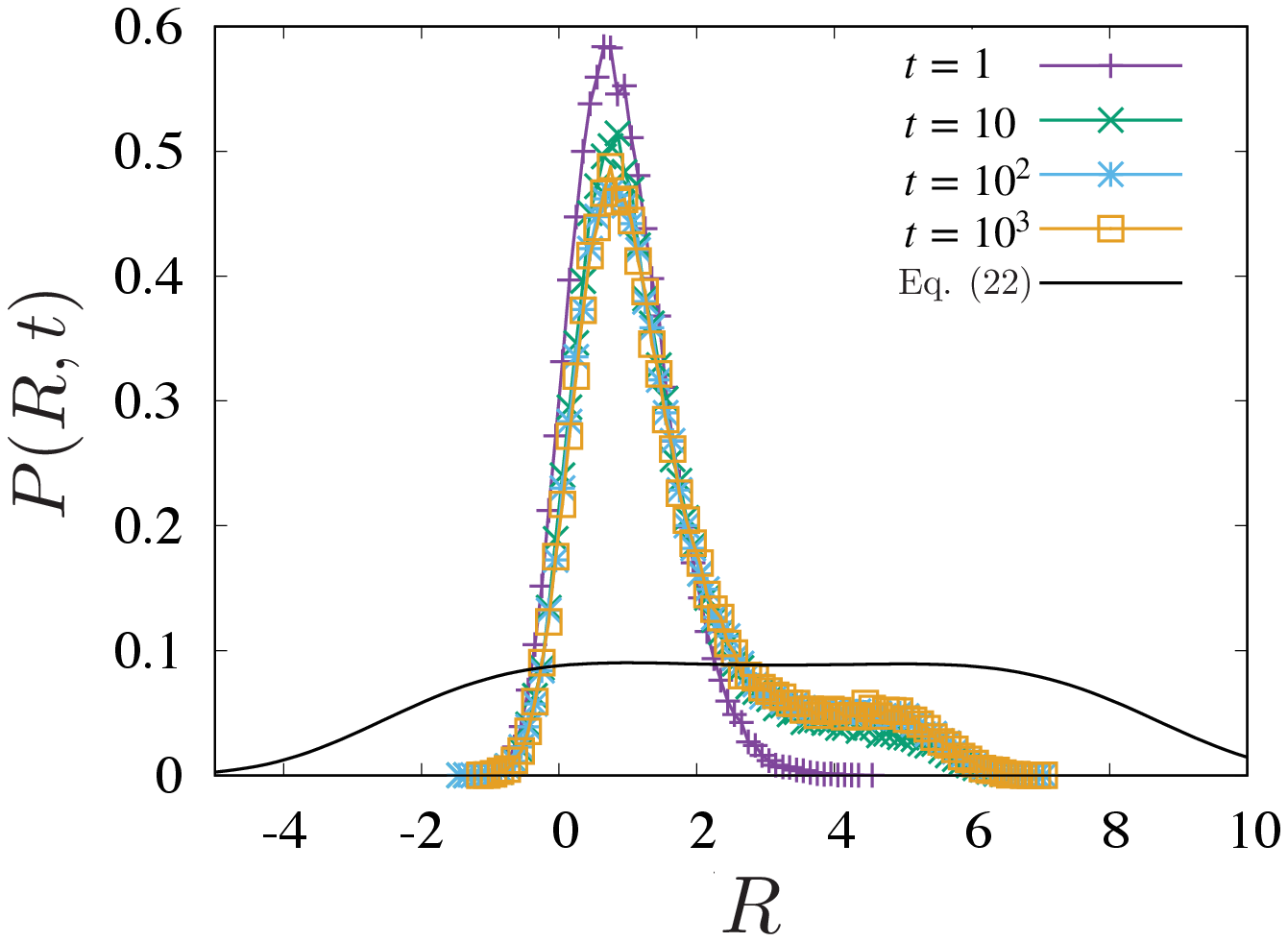}
\caption{Propagators of the GLE $R(t)$ for different times, where the memory kernel is a power-law form  
($\alpha =0.5$), the initial condition is $P(R,0) = \delta(R-1)$ and  the parameters of the double-well potential are the same as those in Fig.~\ref{trajectory}. 
Symbols with lines are the results of numerical simulations. 
The solid line shows the equilibrium distribution, i.e., Eq.~(\ref{boltzmann dist}). }
\label{propagators}
\end{figure}

We discuss how the non-Markovian fluctuating diffusivity affects the global diffusivity. Since the fluctuating diffusivity $D(t)$ is a stationary process, 
the stationary distribution exists. Therefore, the global diffusivity is described as 
\begin{equation}
\langle D \rangle_{\rm st} = \int_{-\infty}^{\infty} D(R) P_{\rm st}(R) dR,
\label{DC stationary}
\end{equation}
where $P_{\rm st}(R)$ is a stationary distribution for $R(t)$. Figure~\ref{propagators} shows the propagators for $R(t)$ when the memory kernel is 
a power-law form with exponent $\alpha=0.5$. The propagator actually converges to a stationary distribution. However, the stationary distribution clearly deviates 
from the Boltzmann distribution, i.e., Eq.~(\ref{boltzmann dist}). There is a trace of the initial condition, i.e.,  $R(0)=1$. In particular, the stationary distribution has 
a peak around the initial condition. We confirm numerically that the stationary distribution crucially depends on the initial condition. 
Therefore, the effect of the initial condition is everlasting. This non-uniqueness of the stationary distribution for $R(t)$ generates the non-Markovian effect of 
the fluctuating diffusivity on the global diffusivity. When the initial condition is the left bottom of the double-well potential, the global diffusivity is greater than that 
for the Markovian fluctuating diffusivity. Furthermore, a strong non-Markovianity increases the global diffusivity.

\section{Conclusion}

We have investigated  how non-Markovian fluctuating diffusivities affect the global diffusivity using the LEFD. 
The MSD exhibits normal diffusion in the LEFD even when the fluctuating diffusivity is non-Markov. 
We have found that non-Markovian fluctuations of fluctuating diffusivity change the global diffusivity, i.e, the diffusion coefficient in the MSD.
This non-Markovian effect is a consequence of the everlasting effect of the stationary distribution  on the initial condition in the fluctuating diffusivity, where 
the dynamics are described by the GLE through the relation $D(t)=1/R(t)$. 
The stationary distribution deviates from the Boltzmann distribution and depends crucially on the initial condition. 
When the dynamics of the gyration radius of a protein are described by the GLE, the LEFD with a Stokes-Einstein-like relation $D(t)=1/R(t)$ is a model 
of the center-of-mass trajectory of the protein. 
Therefore, our results indicate that non-Markovian conformational fluctuations play an important role in regulating the diffusivity of the protein.

\section*{Acknowledgement}
The authors thank Tomoshige Miyaguchi for fruitful discussions on numerical simulations of the GLE. 
T.A. was supported by JSPS Grant-in-Aid for Scientific Research (No.~C 21K033920).



%

\end{document}